\documentclass[twocolumn]{autart}    % Enable this line and disable the 
\usepackage{graphicx} 
\usepackage{cite}
\usepackage{amsmath,amssymb,amsfonts}
\usepackage{graphicx}
\usepackage{subcaption}
\usepackage{textcomp}
\usepackage{mathtools}
\usepackage{xcolor}
\usepackage{fourier}
\usepackage{epstopdf}
\usepackage{hyperref}
\usepackage{float}

\newtheorem{theorem}{\textbf{Theorem}}

\newtheorem{corollary}{\textbf{Corollary}}

\newtheorem{remark}{\textbf{Remark}}
\newenvironment{proof}{{\emph{\textbf{Proof:}} }}{\hfill $\square$}

\begin{document}

\begin{frontmatter}

\title{External Bias and Opinion Clustering in Cooperative Networks\thanksref{footnoteinfo}} % Title, preferably not more 
\thanks[footnoteinfo]{This paper was not presented at any IFAC meeting. Corresponding author T.~Tripathy. The first three authors have contributed equally towards the manuscript.}
%\thanks[footnoteinfo]{This work was not supported by any organisartion.}

\author{Akshay Nagesh Kamthe}\ead{ankamthe21@iitk.ac.in},    % Add the 
\author{Vishnudatta Thota}\ead{thotav@iitk.ac.in},               % e-mail address 
\author{Aashi Shrinate}\ead{aashis21@iitk.ac.in}, % (ead) as shown
\author{Twinkle Tripathy}\ead{ttripathy@iitk.ac.in} 

%\address[a1]{Graduate student}  % Please supply                                              
%\address[a2]{Undergraduate student}             % full addresses
%\address[a3]{Graduate student}        % here.
\address{Department of Electrical Engineering, Indian Institute of Technology Kanpur, Uttar Pradesh, India 208016.} 
          
                             % keyword list or with the 
                                          % help of the Automatica 
                                          % keyword wizard

%Further, in the given framework, we prove the existence of trajectories of any length $l_o\in(l_m,\infty)$. By choosing suitable values of initial conditions, it is also possible to guarantee the existence of curvature bounded trajectories of any arbitrary lengths.

\begin{abstract}                          % Abstract of not more than 200 words.
In this work, we consider a group of $n$ agents which interact with each other in a cooperative framework. A Laplacian-based model is proposed to govern the evolution of opinions in the group when the agents are subjected to external biases like agents' traits, news, \textit{etc}. The objective of the paper is to design a control input which leads to any desired opinion clustering even in the presence of external bias factors. Further, we also determine the conditions which ensure the reachability to any arbitrary opinion states. Note that all of these results hold for any kind of graph structure. Finally, some numerical simulations are discussed to validate these results.
\end{abstract}
%the existence of a non-empty subset of the feasible trajectories which have a specified desired length ... By choosing suitable values of initial conditions, it is also possible to guarantee the existence of curvature bounded trajectories of any arbitrary lengths. ..... The notion of curvature boundedness on such trajectories is introduced and the reachability of lengths for such trajectories is analysed.
\end{frontmatter}
\section{Introduction}
Opinion dynamics in networked environments have captured the interest of diverse fields for a considerable time. It finds applications in various domains like analysis of voting patterns \cite{negativebias2017}, prediction of social media trends \cite{gorodnichenko2021social}, collective animal behaviours \cite{srivastava2017bio}, \textit{etc}. The focus of this paper is on \textit{opinion clustering} in which more than two clusters of agents' opinions are eventually formed in the network. Arbitrary clustering occurs in a signed network in \cite{trust_mistrust2016} when opinions evolve by the DeGroot-based model and the subnetwork of globally reachable nodes is structurally balanced. Other extensions of DeGroot's model in the literature include the homophily-based Hegselmann-Krause model \cite{HK2002}, the biased assimilation model \cite{Dandekar2013biasedassim}, the confirmation bias model \cite{Yanbing2018confbias} and scaled consensus \cite{ROY2015259} that explain opinion clustering. In these works, the clusters depends on the initial conditions and the network topology, making it difficult to reach a desired opinion cluster. 

 Recent works {\cite{xia2011clustering,qin2013cluster,9388882,de2022consensus,tomaselli2023multiconsensus,MA2024105699} have addressed the opinion clustering problem towards applications such as task allocation, rendezvous problems, \textit{etc}. In \cite{xia2011clustering,qin2013cluster,9388882}, the authors propose a Laplacian-based pinning control law for a network partitioned into sub-groups of agents which satisfy the \textit{in-degree balance condition}. To promote opinion separation between agents in different subgroups, signed interactions are introduced between the latter; the agents within a sub-group are constrained to have strong non-negative couplings. With similar constraints, the agents form k-partite consensus in \cite{de2022consensus} in an undirected network wherein those within each sub-group cooperate, and others necessarily compete. The authors in \cite{MA2024105699} present topology-based conditions for clustering in weakly connected digraphs using Laplacian flows.
 The works in \cite{qin2013cluster} and \cite{9388882} explore the network topologies that allow desired clustering in the absence of strong interactions among agents within a subgroup. Using another approach in \cite{tomaselli2023multiconsensus}, the authors propose to partition the agents into subgroups based on graph symmetries in connected undirected networks and design a control law to drive them to the subspace of the dominant eigenvector of graph adjacency matrix for any desired opinion clustering.

 % by converging to a desired form of the dominant eigenvector.}%the authors propose to partition the agents into subgroups based on graph symmetries; any desired opinion clustering is achieved in a connected undirected network by converging to a desired form of the dominant eigenvector.}%A previous work \cite{gambuzza2020controlling} assigns any graph symmetries by modifying the network topology;

In the area of opinion dynamics, along with inter-agent interactions, an agent can also be influenced by individual prejudices \cite{varma2020analysis}, external sources such as news and social media \cite{vicaro2016misinfo}, \textit{etc}.  The Taylor's model\cite{taylor1968towards} is a continuous-time model that considers the evolution of opinions in the presence of agents' biases. The effect of agent's biases is further explored in \cite{baumann2020laplacian}, and a measure to quantify a stubborn agent's influence depending on its position in a network is proposed. In \cite{friedkin1990social}, the authors propose the Friedkin-Johnsen (FJ) model which is a discrete-time analogue of the Taylor's model. A topology-based partitioning of the network is proposed in \cite{YAO2022105267} under the FJ framework along with the criteria to achieve cluster consensus under the influence of single and multiple stubborn agents. News and social media are also frequently leveraged to shape opinions in socio-political scenarios. The effect of misinformation and rumors on opinion formation is explained in \cite{vicaro2016misinfo}. In \cite{fedele2023impact}, the authors determine the impact of factors like news, media, and political leaders on opinion formation using a simplified Taylor's model; the proposed approach provides an accurate estimation of the final opinions.
%a measure is derived to denote how efficiently a single stubborn agent affects the opinions of a group in consensus with opinion evolving by the Taylor's model. Further, the steady-state opinions of agents whose opinions evolve in the presence of two hostile biased agents are determined.}
%Along with inter-agent interactions, external factors can also influence the opinions of agents \cite{varma2020analysis}. For example, external influences such as news and social media are frequently leveraged to shape opinions in socio-political scenarios. The effect of spreading misinformation and rumours on opinion formation is explained in \cite{vicaro2016misinfo}. In \cite{simple2019}, a simple model with a Monte Carlo approach has been used to explore how people form opinions and vote in a two-party system where the population is subjected to an external bias that favours the minority. The results presented in \cite{negativebias2017} illustrate the influence of external bias intensity on the outcome of a two-party election.% The prevailing literature primarily examines the influence of external biases, particularly within two-party voting scenarios. 

In contrast to the works discussed so far, the current work proposes an approach to achieve opinion clustering within a cooperative network in the presence of external bias factors. The opinion evolution is governed by a Laplacian-based model appended with a constant \textit{bias term} and a control input. The bias term represents the cumulative effect of age, socio-economic conditions of the agents, social media, \textit{etc}. In this work, we extend the analysis in \cite{fedele2023impact} to directed networks and explore all the potential outcomes of opinion evolution in the given framework for any value of the bias term and the initial conditions. Contrary to the works in \cite{MA2024105699,YAO2022105267}, we present a methodology to design the control input to achieve any desired opinion clustering despite the presence of external bias. The major advantage of the work is that the analysis presented holds for any arbitrary network topology, unlike the works \cite{xia2011clustering,qin2013cluster,9388882,de2022consensus,tomaselli2023multiconsensus,YAO2022105267,MA2024105699} which widens its scope of applicability.
%In contrast to the works discussed so far, the current work explores the potential outcomes of opinion evolution when subjected to external biases within a network of arbitrary graph connectivity. To do so, we propose using a Laplacian-based model appended with a constant \textit{bias term}. The latter represents the cumulative effect of socio-economic conditions of the agents, their professional and familial roles, news, media, \textit{etc}. In the given framework, we determine all possible kinds of emergent behaviours. Thereafter, we also present the conditions required to achieve any desired opinion pattern or clustering. The motivation stems from the fact that a desired opinion pattern can be used to mitigate the adverse effects of external bias factors like polarisation. Additionally, the analysis presented in this work holds for any arbitrary graph structure which widens the scope of applicability of the work.

The paper is organized as follows: Section \ref{section:two} contains some necessary preliminaries from graph theory. Section \ref{opinionmodel} presents the model which governs the evolution of opinions. The effects of external bias factors and the conditions required to achieve any desired clustering of opinions are discussed in Sections \ref{The effect of bias on opinion formation} and \ref{desired_clustering}, respectively. Section \ref{section:four} demonstrates these results through numerical simulations. To further elaborate on the results, a realistic example has been discussed in Section \ref{sec:discussion}. Finally, Section \ref{sec:conclusion} concludes the paper.% with some insights into the possible future research directions.
\section{Preliminaries}
\label{section:two}
A weighted graph is represented by $\mathcal{G}=(\mathcal{V},\mathcal{E})$ where $\mathcal{V}=\{1,2,\cdots,n\}$ is the set of nodes, $\mathcal{E} \subseteq \mathcal{V} \times \mathcal{V}$ is the set of edges of the graph $\mathcal{G}$. The nodes and the edges represent the agents and their interactions in the multi-agent framework, respectively. The edge $(i, j) \ \in \mathcal{E} $ denotes an edge from node $i$ to node $j$. In an undirected graph, nodes $i$ and $j$ are neighbours if there exists an edge $(i,j) \in \mathcal{E}$. In an directed graph, for an edge $(i,j)$, the node $i$ is in-neighbor of $j$, and node $j$ is out-neighbor of node $i$. The neighbourhood of a node $i$ is $N_{i}= \{j \in \mathcal{V}$ $|$ $(i,j) \ \in \mathcal{E} \}$.

The Adjacency matrix $A \in \mathbb{R}^{n\text{x}n}$ for the graph $\mathcal{G}$ is denoted by $A=\{a_{ij}\}$. The entry $a_{ij}$ is the weight of the edge $(i, j)$. The degree of a node refers to the number of neighbours for undirected graphs. For digraphs, the notions of in-degree and out-degree exist, which refer to the number of in-neighbors and out-neighbors of a node. The out-degree matrix $D=\{d_{ij}\}$ is defined as $d_{ij}=\sum_{k \in N_i}a_{ik}$ for $i=j$ and $0$ otherwise. $\mathbf{1}_n$ and $\mathbf{0}_n$ denote column vectors with all entries equal to $+1$ and $0$, respectively. The matrix $I_n$ denotes the identity matrix of dimension $n$. The Laplacian matrix $L \in   \mathbb{R}^{n\text{x}n}$ for the graph $\mathcal{G}$ is defined as 
\begin{equation} \label{eq:laplacian}
L=D-A.
\end{equation}
It follows from eqn. \eqref{eq:laplacian}, that $L \mathbf{1}_n= \mathbf{0}_n$ therefore, Laplacian matrix $L$ will always have a zero eigenvalue in a cooperative framework.
The non-zero eigenvalues of the Laplacian matrix have a strictly positive real part.

Next, we discuss some graph properties. A sink is a node with a zero out-degree. A node that is reachable from every other node is called a globally reachable node. If every node is reachable from every other node, the graph is \textit{strongly connected}. A weakly connected graph is a directed graph that is not strongly connected, but its undirected version is connected. A condensation graph $C(\mathcal{G})$ of a digraph $\mathcal{G}$ has nodes consisting of strongly connected components of the graph $\mathcal{G}$. There exists an edge from node $I$ to node $J$ in $C(\mathcal{G})$, if and only if there exists an edge from a node in $I$ to a node in $J$ in graph $\mathcal{G}$ where $ I$ and $J$ denote the strongly connected components of graph $\mathcal{G}$ in $C(\mathcal{G})$. %The condensation graph of a weakly connected graph is weakly connected (see Lemma 3.2 in \cite{FB-LNS}).
%
%\begin{definition}
% A sink is a node with a zero out-degree. A set of strongly connected nodes in a graph $\mathcal{G}$ also behaves as a sink if it forms a node with $0$ out-degree in its condensation graph $C(\mathcal{G})$. 
%\end{definition}

\section{Opinion Modelling}
\label{opinionmodel}
In this work, we study the evolution of opinions in a group of $n$  agents interacting in a cooperative social framework. The evolution of opinions, in general, depends on various factors which are broadly classified as endogenous and exogenous %with respect to the group 
\cite{Noah1997}. The endogenous factors arise from the interpersonal relationship between the agents. The exogenous factors are external to the group and depend on social factors like gender, age, socio-economic conditions, and media.
We categorize the exogenous factors into two types: (a) those which are pre-existing and cannot be modified or controlled, viz., age and gender, denoted as a \textit{bias term} $\mathbf{b}$ (b) those which can be controlled, viz., news and advertisements, denoted as a \textit{control input} $\mathbf{u}$. Along with interpersonal relations, the opinions of the agents are also affected by these exogenous factors, given cumulatively by $\mathbf{b}+\mathbf{u}$. We treat $\mathbf{u}$ as the control input, which is used to achieve the desired opinion patterns like consensus, polarisation, and clustering. 

%A real world example of the effect of exogenous factors on opinion evolution in a cooperative network can be observed in the Reddit communities. 
%\textit{Example:} When users join the Reddit platform to discuss their topics of mutual interests, such as technology, sports, movies, \textit{etc.}, they are cooperative with each other. The users come from various walks of life, implying that they have their own set of specific interests or biases on some of the (sub)topics. Owing to this, with time, the users formed subgroups (clusters) within the group. For example, people eventually can form sub-groups of `technology' like programming, hardware, \textit{etc}. The `interest' in the subtopics can be viewed as exogenous factors, which lead to the formation of clusters within the groups even in the presence of cooperation.
 \textit{Example \cite{glasper2021reducing}:} During the COVID-19 pandemic, the population of the US was polarised into pro-vaccine and anti-vaccine groups due to the influence of their existing beliefs, misinformation spread by social media interactions and bots, \textit{etc.} All of these are categorised as a bias term \textbf{b}. The Government and health organisations used advisories and fact checks to counter the misinformation on social media. Further, they encouraged well-known figures from various fields to get immunised and raise awareness among people. These targeted interventions can be categorised under control input \textbf{u}.

%Throughout this work, we consider the interactions between agents to be cooperative, implying that the edge weights of the graph $\mathcal{G}$ are positive. 
Considering the presence of both endogenous and exogenous factors, in this work, we propose the following continuous time  opinion model for the $i^{th}$ agent,
\begin{equation} \label{eq:opinion_model}
\dot{x}_i=\sum_{j \in N_i}a_{ij}(x_j(t)-x_i(t))+b_i+u_i
\end{equation}
where $x_i$, $b_i$, and $u_i$ represent the opinion, the bias term, and the control input of the $i^{th}$ agent, respectively. In eqn. \eqref{eq:opinion_model}, the term $a_{ij}$ represents agent $j$'s influence on agent $i$. In vector form, the opinion dynamics given in eqn. \eqref{eq:opinion_model} can be re-written as,
\begin{equation} \label{eq:linear_system}
\dot{x}=-Lx+\mathbf{b}+\mathbf{u}
\end{equation}
where $L$ is the Laplacian matrix for the underlying graph $\mathcal{G}$ as defined in eqn. \eqref{eq:laplacian}, $\mathbf{b}=(b_1,b_2,\cdots,b_n)^T$ is the constant bias vector and $\mathbf{u}=(u_1,u_2,\cdots,u_n)^T$ is the control input vector. This model is used to reflect the effect of the societal bias factors $b_i$ on the evolution of opinions. 
In real world scenarios like bimodal coalitions, duo-polistic markets, and competing international alliances, polarisation can be extremely undesirable. Hence, we further aim to design $u_i$ so as to mitigate the undesired effects of $b_i$ through the desired collective behaviour of clustering.

\section{The effect of bias on opinion formation}
\label{The effect of bias on opinion formation}
In this section, we study the effect of pre-existing exogenous bias factors on opinion formation in the absence of any control input $\mathbf{u}$. Without the bias vector $\mathbf{b}$, the opinion model given in eqn. \eqref{eq:opinion_model} becomes the well-studied Laplacian flow \cite{FB-LNS} which leads to consensus for several graph structures; the additional term $\textbf{b}$ can sway opinion states away from consensus.

To study the evolution of the opinion states with time, we rewrite $-L$ using its canonical decomposition as $-L=VJW^{T}$ where $V$ and $W$ are the matrices consisting of the right eigenvectors $\mathbf{v}_i$ and the left eigenvectors $\mathbf{w}_i$ of $-L$, respectively, along their columns for $i \in \{1,...,n\}$. The matrix $J$ is the block diagonal Jordan normal form (see Section 2.1.2 in \cite{FB-LNS}). 

The spectrum of $-L$ is denoted by $\sigma=\{\sigma_1, \sigma_2,\cdots,\sigma_n\}$. Without loss of generality, the initial time $t_0$ is assumed to be $0$ throughout the paper. Then, the solution of eqn. \eqref{eq:linear_system} with $\mathbf{u}=\mathbf{0}_n$ \cite{chen1984linear} can be written as,
\begin{equation} \label{eq:values_nosink}
x(t) = Ve^{Jt}W^{T}x_0+ V\int_{0}^te^{J(t-\tau)}d\tau W^{T}\mathbf{b}.
\end{equation}
The subsequent result aids our understanding of how biases affect opinion formation, with a discussion on the stability aspects of the arising collective behaviours.
\begin{theorem}\label{Effectofbias_undirected}
The system \eqref{eq:linear_system} with $\mathbf{u}=\mathbf{0}_n$ admits a stable solution regardless of the connectivity of the graph, if and only if the following equation holds \begin{equation}\label{stability_condition}\mathbf{w}_i^T\mathbf{b}=0 \ \forall \sigma_i=0\end{equation}
where $i\in\{1,2,\cdots,n\}. $ Otherwise, it is unstable. 
Let $n_z$ be the number of zero eigenvalues of $-L$, then for the stable case, at steady state $\lim_{t\to\infty}x(t)$ can be given as,
\begin{equation}\label{eq:convergence2}
\bar x=V\left[\mathbf{w}_1^Tx_0,\cdots,\mathbf{w}_{n_z}^Tx_0,\left[-(\Tilde{J}^{-1})\tilde{W}^T\mathbf{b}\right]^T\right]^T
\end{equation}
where $\tilde{J}$ and $\tilde{W}$ are submatrices of $J$ and $W$, respectively, pertaining to the non-zero eigenvalues of $-L$.
For the unstable case, the weighted average of the opinion states evolves with time, as given below, 
\begin{align} 
\label{eq:constant_nonzero}
\mathbf{w}_i^{T}x(t)=\mathbf{w}_i^{T}(x_{0}+\mathbf{b}t)  \qquad \forall \sigma_i=0,t\geq 0. 
\end{align}
\end{theorem}
\begin{proof}
We start the proof by showing that eqn. \eqref{stability_condition} is necessary for convergence irrespective of network connectivity. It is known that the Jordan blocks of the Laplacian matrix with zero eigenvalues are of size $1$, and the multiplicity of zero eigenvalues depends on the connectivity of the graph \cite{FB-LNS}.

We decompose $J$ into two parts, one with only zero eigenvalues and the other with non-zero values $\tilde{J}$, which is invertible. Hence, we can rewrite eqn. \eqref{eq:values_nosink} as, 
\begin{equation}
\label{timevarying_x(t)}
x(t)=Ve^{Jt}W^{T}x_0+V\begin{bmatrix}
C &0\\
0 &(e^{\tilde{J}t}-I)\tilde{J}^{-1}
\end{bmatrix}W^{T}\mathbf{b}
\end{equation}
where $C =diag(t,t,\cdots,t)$ is a diagonal matrix with the time-varying terms occurring due to the integration of the part of $J$ corresponding to the zero eigenvalues, the matrix $C$ is a square matrix with dimension $n_z$.
\begin{align*}
\lim_{t\to\infty}x(t) = \sum_{i=1}^{n_z}\mathbf{v}_{i}\mathbf{w}_{i}^{T}x_0+& \lim_{t\to\infty}V\Big[
\mathbf{w}_1^T\mathbf{b}t, \cdots, \mathbf{w}_{n_z}^T\mathbf{b}t,\\
& \big[(e^{\tilde{J}t}-I)\Tilde{J}^{-1}\tilde{W}^T\mathbf{b} 
\big]^T\Big]^T.
\end{align*}
The solution $x(t)$ becomes stable when the time-varying terms do not exist. By applying eqn. \eqref{stability_condition}, the time-varying terms vanish such that
\[
\bar x = \sum_{i=1}^{n_{z}}\mathbf{v}_{i}\mathbf{w}_{i}^{T}x_0+ V\left[
0,\cdots,0,
\left[-\Tilde{J}^{-1}\Tilde{W}^{T}\mathbf{b}\right]^T
\right]^T
\]
% \begin{equation}
% \label{eq:x_tilda_final_state_equation}
% \bar x = \sum_{i=1}^{n_{z}}\mathbf{v}_{i}\mathbf{w}_{i}^{T}x_0+ V\left[
% 0,\cdots,0,
% {\color{blue}\left[-\Tilde{J}^{-1}\Tilde{W}^{T}\mathbf{b}\right]^T}
% \right]^T
% \end{equation}
% \[
% \bar x = \sum_{i=1}^{n_{z}}\mathbf{v}_{i}\mathbf{w}_{i}^{T}x_0+ V\left[
% 0,\cdots,0,
% -\Tilde{J}^{-1}\Tilde{W}^{T}\mathbf{b}
% \right]^T.
% \]
Further rearrangement results in a stable solution  as given by eqn. \eqref{eq:convergence2}.

Next, we consider the case when eqn. \eqref{eq:linear_system} admits a stable solution. At steady state, we have $\dot{x}=0$ which implies $-L\bar x+\mathbf{b}=0$. Rearranging and premultiplying it with $\mathbf{w}_i^T$, we get
$\mathbf{w}_i^T\mathbf{b}=\mathbf{w}_i^TL\bar x$. Note that $\mathbf{w}_i^TL=\mathbf{0}_n^T \ \forall \ \sigma_i=0$ as since $\mathbf{w}_i$ is the left eigenvector corresponding to the $\sigma_i=0$. This implies $\mathbf{w}_i^T\mathbf{b}=0$ for every zero-eigenvalue. This concludes the discussion on the stability of system \eqref{eq:linear_system}.

In the case when system \eqref{eq:linear_system} is unstable, $\mathbf{w}^T_i\mathbf{b}\neq 0$. Then, $x(t)$ can be obtained by simplifying eqn. \eqref{timevarying_x(t)} as,
\begin{multline}
\label{diverge}
 x(t)=V\bigg[\mathbf{w}_1^T(x_0+\mathbf{b}t),\cdots,\mathbf{w}_{n_z}^T(x_0+\mathbf{b}t),  
 \left[-(\Tilde{J}^{-1})\tilde{W}^T\mathbf{b}\right]^T\bigg]^T    
\end{multline}
The eigenvectors of $L$ are normalized to satisfy $W^TV=I_n$. Hence, the terms with $\mathbf{v}_{j\neq i}$, $j\in\{1,2,\cdots, n\}$, disappear resulting in eqn. \eqref{eq:constant_nonzero}. Hence, proved. 
\end{proof}

The stable solution given in eqn. \eqref{eq:convergence2} simply means that the opinion states converge to finite values. Theorem \ref{Effectofbias_undirected} shows that such a behaviour can be achieved irrespective of the graph connectivity. The steady state solution given in eqn. \eqref{eq:convergence2} can be refined further for some specific graph structures whose Laplacian matrices have zero as a simple eigenvalue. 
\begin{corollary}
\label{lem:undirected}
For a connected undirected graph, system \eqref{eq:linear_system} admits a stable solution if and only if $\sum_{i=0}^{n}b_{i}=0$. At steady state, eqn. \eqref{eq:convergence2} becomes,
\begin{equation}\label{convergence2_undir}
\bar x=V\left[\sum_{i=1}^{n}\frac{x_{0i}}{n},\frac{-\mathbf{w}_2^T\mathbf{b}}{\sigma_2},\frac{-\mathbf{w}_3^T\mathbf{b}}{\sigma_3},\cdots,\frac{-\mathbf{w}_n^T\mathbf{b}}{\sigma_n}\right]^T
\end{equation}
For $\sum_{i=1}^{n}b_{i} \neq0$ system \eqref{eq:linear_system} is unstable, and the average value of the opinion states evolves with time as given below,
\begin{equation} \label{eq:constant_nonzero_c_undir}
\frac{1}{n}\sum_{i=1}^nx_{i}(t)=\frac{1}{n}\sum_{i=1}^{n}x_{0i}+ \frac{t}{n}\sum_{i=1}^{n}b_{i}
 \ \ \ t\geq 0\end{equation}
\end{corollary}
\begin{proof}
It is already known that for a connected undirected graph, the Laplacian matrix is symmetric and has a simple $0$ eigenvalue. The corresponding right and left eigenvectors are $\mathbf{1}_n$ and $\mathbf{1}_n / n$, chosen to satisfy $\mathbf{v}_1^T\mathbf{w}_1=1$. Then, eqn. \eqref{stability_condition} simplifies to $\sum_{i=0}^{n}b_{i}=0$ and eqn. \eqref{eq:constant_nonzero} 
to eqn. \eqref{eq:constant_nonzero_c_undir}. Since the inverse of the $i^{th}$ Jordan block $J_i^{-1}=\frac{1}{\sigma_i}$ for all $\sigma_i\neq 0$, eqn. \eqref{eq:convergence2} reduces to eqn. \eqref{convergence2_undir}. Hence, proved.
\end{proof}
\begin{corollary}\label{GRN}
For a digraph containing a globally reachable node(s), system \eqref{eq:linear_system} admits a stable solution if and only if
\begin{align}\label{stability condition dir}
  \sum_{i \in N_G}\mathbf{w}_{1i}b_i=0. 
\end{align}
At steady state, eqn. \eqref{eq:convergence2} becomes,
\begin{equation}\label{directed_general_eqn}
\bar x=V\left[\mathbf{w}_1^Tx_0,\left[-(\Tilde{J}^{-1})\tilde{W}^T\mathbf{b}\right]^T\right]^T
\end{equation}
 where $\mathbf{w}_1=[w_{11},w_{12} is ,\cdots,w_{1n}]^T$ is the left eigenvector and $N_G$ is the set of globally reachable node(s).
For $\mathbf{w}_1^T\mathbf{b} \neq 0$, system \eqref{eq:linear_system} becomes unstable, and the weighted average of the opinion states evolves with time as
\begin{equation} \label{eq:constant_nonzero_c_dir}
\mathbf{w}_1^Tx(t)=\mathbf{w}_1^Tx_{0}+\mathbf{w}_1^T\mathbf{b}t
\end{equation}
\end{corollary}
\begin{proof}
We know that $L$ has a simple zero-eigenvalue (see Theorem $6.6$ in \cite{FB-LNS}). The right eigenvector is $\mathbf{1}_n$ and, for the left eigenvector, $\mathbf{w}_{1i}>0$ $\forall$ $i \in N_G$ and is zero otherwise. Using these results, the rest of the proof follows in the same manner as that of Corollary \ref{lem:undirected}.
\end{proof}
\par Note that strongly connected digraphs form a special case of the graphs discussed in Corollary \ref{GRN} with $N_G=\{1,2,\cdots,n\}$. So far, we have discussed the conditions on the external bias factors $\mathbf{b}$ in Theorem \ref{Effectofbias_undirected}, which are required for stability. However, $\mathbf{b}$ may not even satisfy it. Even if it does, the stable opinion states of the group could become undesirable (e.g. polarisation). In the next section, we present the conditions which guarantee the arbitrary clustering of opinion states while maintaining stability.
\section{Clustering of opinion states}\label{desired_clustering}
In this section, we design the control input $\mathbf u$ to drive the opinion states of system \eqref{eq:linear_system} to desired opinion clusters in the presence of a constant bias vector $\mathbf{b}$. We begin by exploring the conditions required for the stability of the system.  Replacing $\mathbf b$ with $\mathbf{(b+u)}$ in the Theorem \ref{Effectofbias_undirected} results in the stability condition given as $\mathbf{w}_i^T(\mathbf{b}+\mathbf{u})=0 \ \forall \ \sigma_i =0$. Furthermore,
\begin{equation}\label{clustering_type_1}
\mathbf{w}_i^Tx(t)=\mathbf{w}_i^Tx(0) \qquad \forall \sigma_i=0,t\geq 0 
\end{equation}
where $i\in\{1,2,\cdots,n\}.$

Now, we discuss two specific classes of graphs that have zero as simple eigenvalues.
\begin{itemize}
\item For a connected undirected graph, it follows that  $\sum_{i=1}^n(b_i+u_i)=0$ for stability. Then, $\sum_{i=1}^nx_i(t)=\sum_{i=1}^nx_i(0)\ t\geq0$.
\item For a digraph containing a globally reachable node, $\mathbf{w}_1^T(\mathbf{b}+\mathbf{u})=0$ must hold for stability. In this case, $\mathbf{w}_1^Tx(t)=\mathbf{w}_1^Tx(0)\ t\geq0$.
\end{itemize}
It is possible to stabilize the system using eqn.\eqref{clustering_type_1}. However, it is not yet clear if we could achieve the desired opinion state as stable behaviour or not. Now, we try to address this issue starting with the following discussion.
\begin{remark}\label{desired_u} Using eqn. \eqref{clustering_type_1} the reachable set of opinion states is defined as 
\begin{align} 
\label{reachable_set}
X_R\vcentcolon = \{x_d\in\mathbb{R}^n|\mathbf{w}_i^Tx_d=\mathbf{w}_i^Tx_0 \ \forall \ \sigma_i=0\}
\end{align}
where $i\in\{1,2,\cdots,n\}$. If the stable desired opinion states $x_d$ belongs to the reachable set $X_R$, then $\dot{x}=0$ at steady state. This gives $-Lx_{d}+(\mathbf{b}+\mathbf{u})=0$. Then, the control input $\mathbf{u}$ to reach an opinion state $x_d$ is given by 
\begin{equation}\label{eq:u_desired_x}
\mathbf{u}=Lx_{d}-\mathbf{b}.
\end{equation}
\end{remark}
%The method to obtain a desired opinion state $x_d$ not belonging to the reachable set $X_R$ is covered in the following theorem.
Remark \ref{desired_u} shows that it is not possible to achieve $x\in\mathbb{R}^n$\textbackslash$X_R$ using eqn. \eqref{eq:opinion_model}. The following result allows us to make the entire $\mathbb{R}^n$ space reachable.%, thereby allowing us to achieve any desired opinion clustering.%The following theorem addresses how to achieve a desired opinion state $x_d$ that does not belong to the reachable set $X_R$.
\begin{theorem}
\label{time_switching_u}
For a given initial state $x_0$ and a constant bias vector $\mathbf{b}$, system \eqref{eq:linear_system} admits any stable desired opinion state $x_d$, irrespective of the connectivity of the graph by designing the control input $\mathbf{u}$ as given below,%the control input $\mathbf{u}$ satisfies the following conditions,
\begin{enumerate}
    \item[(a)] if $\mathbf{w}_i^Tx_d=\mathbf{w}_i^Tx_0$, then $\mathbf{u}$ is given by eqn. \eqref{eq:u_desired_x},
    \item[(b)] else, $\mathbf{u}$ satisfies the condition,
\begin{align}
\mathbf{w}_i^T(\mathbf{b}+\mathbf{u})=
%\begin{cases}
    (\mathbf{w}_i^Tx_d -  \mathbf{w}_i^Tx_0)/\bar t,~~\forall\sigma_i=0
    \label{t_greater_than_tbar}
    \end{align}
    for $t\leqslant\bar t$ and is given by eqn. \eqref{eq:u_desired_x} for $t>\bar t$ where $\bar t$ is chosen arbitrarily in $(0,\infty)$ and 
$i\in\{1,2,\cdots, n\}$. 
\end{enumerate}
\end{theorem}
\begin{proof}   
When the desired opinion state $x_d$ lies in the reachable set $X_R$, then $\mathbf{u}$ can be calculated using Remark \ref{desired_u}, which leads to condition $(a)$ of the theorem. 

When $x_d\not\in X_R$, it is not possible to reach $x_d$ using the control input given in eqn. \eqref{eq:u_desired_x}. To alleviate this issue, we propose a two-stage solution: first, we reach some $\tilde{x}\neq\{x_0,x_d\}$ from where $x_d$ is reachable; thereafter, we use eqn. \eqref{eq:u_desired_x} to calculate the suitable control input which is required to reach $x_d$. 

Note that it is not possible to reach $\tilde{x}$ starting from $x_0$ through a stable system behaviour as $\mathbf{w}_i^T\tilde{x}\neq\mathbf{w}_i^Tx_0$. In the presence of $\mathbf{b}$ and $\mathbf{u}$, an unstable behaviour can be modelled using eqn. \eqref{eq:constant_nonzero} as,
\begin{align}
\label{eq:unstable_clustering}
\mathbf{w}_i^{T}x(t)=\mathbf{w}_i^{T}(x_{0}+\mathbf{(b+u)}t)    
\end{align}
for $t\leqslant \bar t$, where $\bar t\in(0,\infty)$ is the time till which the unstable behaviour exists such that $\tilde x=x(\bar t)$. 

In the second stage, $\mathbf{w}_i^Tx_d=\mathbf{w}_i^Tx(\bar{t})$ holds as $x_d$ is reachable from $x(\bar{t})$. Then, at $t=\bar t$, eqn. \eqref{eq:unstable_clustering} becomes $\mathbf{w}_i^{T}x_d=\mathbf{w}_i^{T}(x_{0}+\mathbf{(b+u)}\bar t$. Re-arranging this equation gives eqn. \eqref{t_greater_than_tbar} which is the necessary condition the control input $\mathbf{u}$ must satisfy. For $t>\bar t$, using eqn. \eqref{eq:u_desired_x} results in the desired behaviour as discussed in Remark \ref{desired_u}. Hence, proved.
\end{proof}

Using the results discussed in Theorem \ref{time_switching_u}, it is possible to obtain any desired opinion clusters which lie in $\mathbb{R}^n$. Clustering of opinions is often a desired outcome as it prevents polarisation. In the next section, we discuss some simulations to illustrate these results.  
\section{Simulation Results}\label{section:four}
\begin{figure}[ht]
    \centering
\includegraphics[scale=0.3]{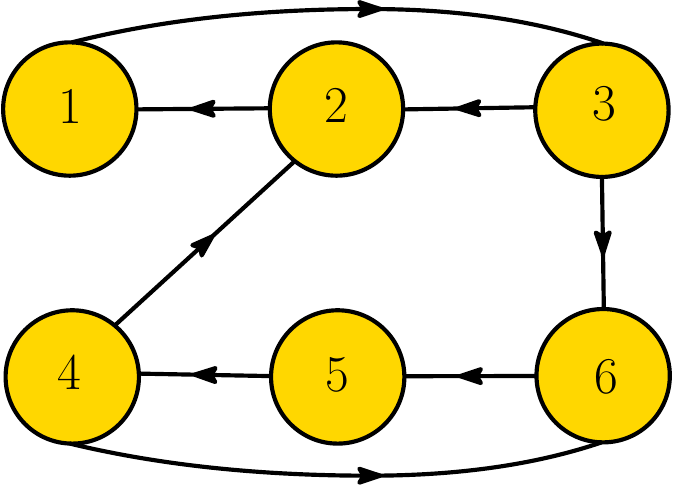}
        \caption{Graph topology}
        \label{Example Graph 1}
        \end{figure}
In this section, numerical simulations are presented to validate the theoretical results discussed in the paper. We consider the graph shown in Fig.   \ref{Example Graph 1}. The initial opinion states are $x(0)=\left[-6, 4, -5, 5, 2, 0\right]^{T}$. Let the external bias factors be such that $\mathbf{b}=\left[-20, 20, 20, -20, 20, -20\right]^{T}$. We know from Theorem \ref{Effectofbias_undirected} that system \eqref{eq:linear_system} admits a stable solution if $\mathbf{b}$ satisfies eqn. \eqref{stability_condition} and $\mathbf{u}=\mathbf{0}_n$. The same can be validated by the evolution of the opinion states shown in Fig. \ref{polarisation}. At steady state, $\bar x = \left[-10,10,10,-10,10,-10\right]^{T}$ which agrees with eqn. \eqref{eq:convergence2}. % Moreover, the weighted average remains the same for all $t\geqslant 0$.
As we can clearly see in Fig. \ref{polarisation}, the bias $\mathbf{b}$ is causing polarisation in the group of agents.
 \begin{figure}[h]
    \centering
 \includegraphics[width=0.45\textwidth]{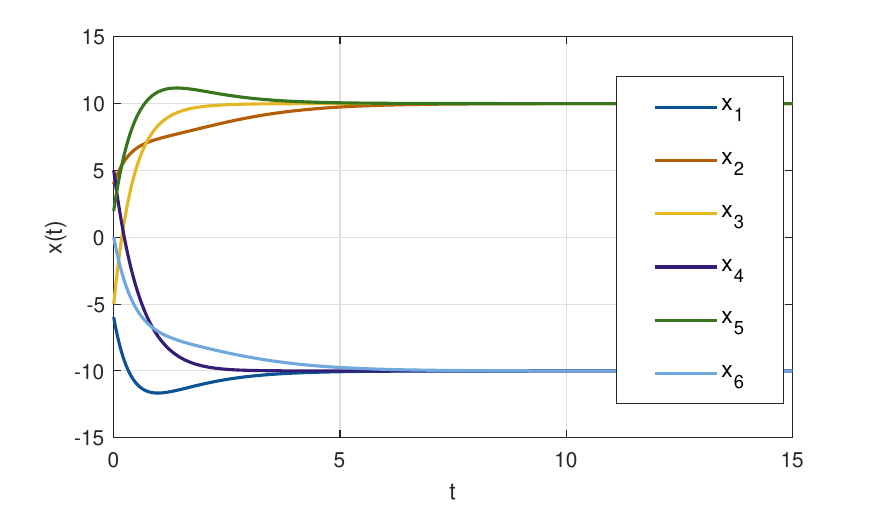}
        \caption{Polarisation of opinion states}
        \label{polarisation}
    \end{figure}
    %\vspace{-0.7cm}
    \begin{figure}[h]
        \centering
 \includegraphics[width=0.45\textwidth]{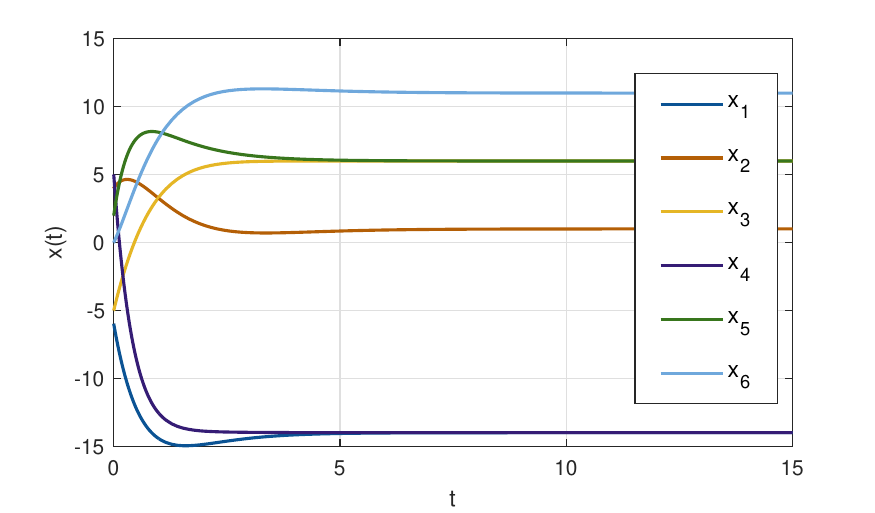}
        \caption{Clustering of opinion states}
        \label{clustering}
    \end{figure}
In this case, an appropriate control input to counter the effect of the bias $\mathbf{b}$ can be designed by using Remark \ref{desired_u} as $\mathbf{u}=\left[0,-5,-2,-20,0,25\right]^{T}$. Then, system \eqref{eq:linear_system} results in the clustering of opinions as shown in Fig. \ref{clustering} where final opinion states are $x_f=[-14,1,6,-14,6,11]^T$. Similarly, any desired opinion pattern can be achieved in an arbitrary graph structure.

\section{Discussion}
\label{sec:discussion}
In the US 2016 Presidential election, it has been shown that social media and fake news had a significant role to play in the outcome of the election \cite{allcott2017social}. In the previous elections due to the absence of such large-scale misinformation campaigns, the individuals were not as polarised. Therefore, it is evident that social media can foster polarising behaviours by using biasing factors as shown in Fig. \ref{polarisation}. By designing the control input using Theorem \ref{time_switching_u}, opinion clustering is achieved as shown in Fig. \ref{clustering}. Thus, the adverse effects of bias can be reduced by designing a suitable control input.
% by making the agents less extreme with respect to each other. %Note that the signs of the control input  %Let us assume an ideal scenario wherein everyone was cooperative to each other. Depending on the social media usage, a subset were subjected to fake news pertaining to one of the parties (positive bias, say), while the others were exposed to fake news pertaining to the other party (negative bias). Fig.\ref{polarisation} shows that such a situation can clearly lead to polarisation which increases hostility in the society. %it can be assumed to have reached almost everyone. The different values

\section{Conclusion}
\label{sec:conclusion}
In this paper, we study the effect of external bias factors on the evolution of opinions in a cooperative network with any arbitrary connectivity. We segregate the external factors into a bias term $\mathbf{b}$ like age, gender, \textit{etc.} which we cannot control or modify, and a control input $\mathbf{u}$ like advertisement, news, \textit{etc.} which we can control. In the presence of a constant external bias $\mathbf b$, we provide the conditions which ensure the stability of the final opinion states of the agents. These conditions guarantee the convergence of the opinion states only to a finite set of values at a steady state. In the proposed framework, we further show that it is possible to design the control input $\mathbf u$ to extend the reachable set of opinion states to $\mathbb{R}^n$. By designing an appropriate control input, the undesirable effects of external bias factors like polarisation can be negated. Furthermore, any desired opinion pattern like consensus, polarisation, and clustering of opinions can also be achieved. In  future, we plan to develop a nonlinear model in the given framework to better emulate a real world scenario.

\bibliographystyle{unsrt}        % Include this if you use bibtex 
\bibliography{main}     

\end{document}